\documentclass[12pt]{article}
\usepackage{epsfig}
\newcommand{\be}{\begin{equation}}
\newcommand{\ee}{\end{equation}}
\newcommand{\ba}{\begin{eqnarray}}
\newcommand{\ea}{\end{eqnarray}}
\newcommand{\se}{\setcounter{equation}{0}}

\newcommand{\1}{^{-1}}
\newcommand{\dg}{^{\dagger}}
\newcommand{\di}{\mbox{d}\,} 
\newcommand{\e}{\mbox{e}}
\newcommand{\ga}{\gamma_5}
\newcommand{\h}{\frac{1}{2}}
\newcommand{\Id}{\mbox{1\hspace{-1.02mm}l}}   
\newcommand{\la}{\lambda}
\newcommand{\mb}[1]{\quad\mbox{ #1 }\quad}
\newcommand{\mo}[1]{^{(\mbox{\tiny #1})}}
\newcommand{\na}[1]{\nabla_{#1}}
\newcommand{\ra}{\rightarrow}
\newcommand{\re}[1]{(\ref{#1})}
\newcommand{\vp}{\varphi} 
\newcommand{\W}{_{\mbox{\tiny W}}} 
\newcommand{\ka}{\kappa} 
\newcommand{\mi}[1]{_{\mbox{\tiny #1}}}
\newcommand{\mii}[2]{_{\mbox{\tiny #1}\,#2}}
\newcommand{\Nh}{\hat{N}}
\newcommand{\Tr}{\mbox{Tr}}

\newcommand{\C}{{\cal C}} 
\newcommand{\sy}{\scriptscriptstyle}

\begin{document}
\renewcommand{\baselinestretch}{1.0} \small\normalsize

\hfill {\sc HU-EP}-02/13

\vspace*{1.2cm}

\begin{center}

{\Large \bf More chiral operators on the lattice} 

\vspace*{0.9cm}

{\bf Werner Kerler}

\vspace*{0.3cm}

{\sl Institut f\"ur Physik, Humboldt-Universit\"at, D-10115 Berlin, 
Germany}
\hspace{3.6mm}

\end{center}

\vspace*{1.2cm}

\begin{abstract}
Instead of the Ginsparg-Wilson (GW) relation we only require generalized 
chiral symmetry and show that this results in a larger class of Dirac 
operators describing massless fermions, which in addition to GW fermions 
and to the ones proposed by Fujikawa includes many more general ones. 
The index turns out to depend solely on a basic unitary operator.  We use 
spectral representations to analyze the new class and to obtain detailed 
properties. We also show that our weaker conditions still lead properly 
to Weyl fermions and to chiral gauge theories. 

\end{abstract}

\vspace*{0.8cm}

\section{Introduction}
 
\hspace{3mm}    
In a large part of the works on chiral fermions the Dirac operator $D$ is
required to satisfy the Ginsparg-Wilson (GW) relation \cite{gi82}
\be
\{\ga,D\}=\rho^{-1}D\ga D \;,
\label{GW}
\ee
with a real constant $\rho$, and to be $\ga$-hermitian,
\be
D\dg=\ga D\ga\;. 
\label{ga5}
\ee
Conditions \re{GW} and \re{ga5} imply that the operator $\Id-\rho\1 D$ is
unitary and $\ga$-hermitian. Therefore, instead of imposing \re{GW} and 
\re{ga5} as is done in such works, it is completely equivalent to require $D$ 
to have the form
\be
D=\rho(\Id-V) \mb{with} V\dg=V^{-1}=\ga V \ga\;.
\label{DN}
\ee

In this notation the condition for the generalized chiral symmetry of the 
classical action of L\"uscher \cite{lu98i} in the form used in Ref.~\cite{lu98} 
reads 
\be
\ga D + D \ga V = 0 \mb{with} V\dg=V^{-1}=\ga V \ga\;.
\label{gg}
\ee
In the present paper we observe that \re{gg} provides a weaker condition than 
the GW relation \re{GW}. Accordingly, instead of \re{GW} and \re{ga5} as are 
usually imposed, we here only require \re{gg} and \re{ga5} and investigate the 
resulting larger class of Dirac operators.

We note that using \re{ga5} one gets from \re{gg} 
\be
D+D\dg V=0 \;,\quad D\dg+DV\dg=0\;,
\label{gg1}
\ee
by which it follows that $V$ and $D$ commute and that $D$ is normal, 
\be
[V,D]=0\;,\quad
DD\dg=D\dg D\;.
\label{COM}
\ee
To account for this we require $D$ to be a function of $V\,$,
\be
D=F(V)\;,
\ee 
i.e.~to depend on $V$ and possibly on constants, however, not on any other 
operator. 

We show that in addition to GW fermions with \re{GW} and to the ones proposed 
by Fujikawa \cite{fu00} and their extension \cite{fu02} the new class includes
many more general ones. We nevertheless find that our weaker conditions still
lead properly to Weyl fermions and to chiral gauge theories. Thus it becomes 
obvious that these conditions constitute the more general principle.

In addition to the general conditions \re{gg} and \re{ga5}, we require that $F$ 
allows for a nonvanishing index of the Dirac operator. Our respective analysis
is based on spectral representations. The index turns out to depend solely on 
the basic unitary operator $V$. We introduce spectral functions which on the 
one hand side allow for a construction, which gives insight into possible forms 
of $F\,$, and on the other side describe the location of the spectrum. We
also give a realization of the unitary operator $V$ related to our general 
construction of the Dirac operator $D\,$.

In Section 2 we obtain fundamental properties of the functions $D=F(V)\,$. 
In Section 3 we discuss the special cases of GW fermions, of the ones proposed
by Fujikawa (bringing $D$ of them also into the form $F(V)\,$) and of the 
extension of the Fujikawa type. In Section 4 we develop realizations of 
subclasses. We first study the one where $F(V)$ can be represented by a power 
series. We then give a more general construction for operators of the class, 
which is based on spectral functions. In a next step we further extend this 
construction applying appropriate functions. We then address the question 
which sets of $V$ are related to the subclasses and give a realization of $V$ 
holding for the general construction of $D\,$. Further, by working out a 
particular example, the considerable freedom in this construction is 
illustrated. In Section 5 we show that our conditions still lead to Weyl 
operators and to chiral gauge theories. Section 6 contains our conclusions.

\section{Properties from spectral representation}\se

\hspace{3mm}    
On the finite lattice $V$ has the spectral representation $V=\sum_kv_kP_k$ with
eigenvalues satisfying $|v_k|=1$ and orthogonal projections $P_k=P_k\dg\,$.
Therefore the functions $F(V)$ can be represented by $F(V)=\sum_kf(v_k)P_k\,$. 

Using $\ga$-hermiticity we get the more detailed form
\be
V=P_1^{(+)}+P_1^{(-)}-P_2^{(+)}-P_2^{(-)}+
\sum_{k\;(0<\vp_k<\pi)}(\e^{i\vp_k} P_k\mo{I}+\e^{-i\vp_k} P_k\mo{II})\;,
\label{specv}
\ee
in which the projections satisfy 
\be
\ga P_j^{(\pm)}=P_j^{(\pm)}\ga= \pm P_j^{(\pm)}\;,\quad \ga P_k\mo{I}=
P_k\mo{II}\ga\;. 
\label{PPg}
\ee
The spectral representation of $D=F(V)$ then becomes 
\ba
D=f(1)(P_1^{(+)}+P_1^{(-)})+f(-1)(P_2^{(+)}+P_2^{(-)})\nonumber\\
+\sum_{k\;(0<\vp_k<\pi)}\Big(f(\e^{i\vp_k})P_k\mo{I}+f(\e^{-i\vp_k})
P_k\mo{II}\Big) \quad,
\label{specd}
\ea
in which $D$ is characterized by the function $f(\e^{i\vp})$. Clearly this
function enters only at the values $\e^{i\vp}=\e^{i\vp_k}\,$. However, 
since we want to define $D$ in general (in particular, for any gauge field 
configuration), we have to specify $f(v)$ for all $v=\e^{i\vp}\,$.

Inserting the general form \re{specd} into \re{ga5} and \re{gg1} we obtain
the conditions 
\be
f(v)^*=f(v^*)\;,
\label{cond*}
\ee
\be
f(v)+f(v)^*v=0\;,
\label{cond+}
\ee
respectively, on the functions $f(v)\,$. These conditions imply 
\be
f(1)=0\;,\qquad f(-1) \;{\rm real}\;.
\ee
The operator form of \re{cond*} is
\be
F(V)\dg=F(V\dg)\;.
\label{cond*o}
\ee

In addition to giving $D$ by \re{specd}, the functions $f(v)$ obviously 
describe the location of its spectrum. For continuous $f(\e^{i\vp})$ the 
eigenvalues of $D$ reside on a closed curve in the complex plane which is 
symmetric to the real axis and meets this axis at zero and at the value 
$f(-1)$ (which is required to be nonzero below).

Denoting the dimensions of the right-handed and of the left-handed eigenspace 
for eigenvalue $\pm1$ of $V$ by $N_{+}(\pm)$ and $N_{-}(\pm)$, respectively,
we have from \re{PPg}
\be
\mbox{Tr}(\ga P_1^{(\pm)})=\pm N_{\pm}(1)\;,\;\;
\mbox{Tr}(\ga P_2^{(\pm)})=\pm N_{\pm}(-1)\;,\;\;
\mbox{Tr}(\ga P_k\mo{I})=\mbox{Tr}(\ga P_k\mo{II})=0\;.
\label{TP}
\ee
With this because of $f(1)=0$, using the resolvent $(D-\zeta\Id)^{-1}\,$, we
obtain the index of $D\,$, 
\be 
\lim_{\zeta\rightarrow 0}\mbox{Tr}\Big(\ga\frac{-\zeta}{D-\zeta\Id} \Big)= 
\left\{\begin{array}{ll} N_+(1)-N_-(1) &\mbox{for }f(-1)\ne0\\
N_+(1)-N_-(1)+N_+(-1)-N_-(-1) &\mbox{for }f(-1)=0 \end{array}\right. \quad,
\label{IND}
\ee
and also find
\be 
\lim_{\zeta\rightarrow 0}\mbox{Tr}\Big(\ga\frac{D}{D-\zeta\Id} \Big)=
\left\{\begin{array}{ll} N_+(-1)-N_-(-1) &\mbox{for }f(-1)\ne0\\
0 &\mbox{for }f(-1)=0 \end{array}\right. \quad.
\label{IND1}
\ee
Adding up \re{IND} and \re{IND1} the sum on the l.h.s.~gets 
$\mbox{Tr}(\ga\Id)=0$ so that in any case
\be  
N_+(1)-N_-(1)+N_+(-1)-N_-(-1)=0\;.
\label{sum}
\ee
Because of \re{IND} and \re{sum}, to admit a nonvanishing index we have to 
impose the condition 
\be
f(-1)\ne0\;.
\label{conda}
\ee

After having \re{conda}, according to \re{sum} to allow for a nonvanishing 
index one has also to require that in addition to $1$ the eigenvalue $-1$ of 
$V$ occurs. The sum rule \re{sum} corresponds to the one found in 
Ref.~\cite{ch98} for the special case of Dirac operators which satisfy the GW 
relation \re{GW}.

Using \re{specv} with \re{TP} we find
\be
\mbox{Tr}(\ga V)=N_{+}(1)-N_{-}(1)-N_{+}(-1)+N_{-}(-1)\;,
\ee
so that with \re{sum} we generally get for the index of the Dirac operators $D$
\be
N_+(1)-N_-(1)=\h\mbox{Tr}(\ga V)\;.
\label{INN}
\ee
Thus it turns out that solely the operator $V$ enters for the whole class.
This generalizes the results obtained in the GW case \cite{ha98,lu98i} and
in the overlap formalism \cite{na93} before.

On the infinite lattice the unitary space, in which the operators act, gets
of infinite dimension and, including limit elements, a Hilbert space. Then
to the spectral representations \re{specv} and \re{specd} the continuous parts
$\int_{-\pi}^{\pi}\e^{i\vp}\,\di E_{\vp}$ and
$\int_{-\pi}^{\pi}f(\e^{i\vp})\,\di E_{\vp}$, respectively, are to be added
in which the projector function $E_{\vp}$ is purely continuous. However, the
relations for the index do not change since the continuous spectrum does not 
contribute to them and because the trace Tr for the expressions of interest
can still be defined \cite{ke02}.

We note that condition \re{sum} reflects a fundamental difference to the case
of the Atiyah-Singer (AS) Dirac operator \cite{at68}, for the nonvanishing 
eigenvalues $\la\mii{AS}{j}$ of which one always has 
\be
\Nh_{+}(\la\mii{AS}{j})-\Nh_{-}(\la\mii{AS}{j})=0\mb{for}\la\mii{AS}{j}\ne0\;,
\label{as1}
\ee
where $\Nh_{+}$ and $\Nh_{-}$ are the dimensions of its right-handed and its 
left-handed eigenspaces, respectively. For {\it zero} eigenvalue of the
AS Dirac operator one gets
\be
\Nh_+(0)-\Nh_-(0)=
\mbox{dim ker }D^{(+)}\mi{AS}-\mbox{dim ker }D^{(+)\dag}\mi{AS}
\ee 
where $D^{(+)}\mi{AS}$ is the related Weyl operator (the AS Dirac operator 
being a composition of $D^{(+)}\mi{AS}$ and $D^{(+)\dag}\mi{AS}=
D^{(-)}\mi{AS}$). Thus to obtain a nonvanishing index there it is inevitable 
to allow the space structure itself to vary and to get chirally asymmetric, 
which implies its dependence on the particular gauge field configuration. In 
contrast to this on the lattice there is no such dependence, which is made 
possible by having \re{sum}.

\section{GW-type special cases}\se

\subsection{Ginsparg-Wilson relation}\se

\hspace{3mm} 
The Dirac operators satisfying the GW relation \re{GW} and \re{ga5} are the
simplest special case of the class defined by \re{gg} and \re{ga5}. For these
operators $F$ has the form \re{DN}. As in the more general cases, given the
function $F(V)\,$, suitable $V$ remain to be determined. Thus looking for 
solutions of the GW relation \re{GW}, from the present point of view means to 
look for the appropriate set of operators $V\,$.

The overlap Dirac operator of Neuberger \cite{ne98} is well-known to be of 
form \re{DN} and to provide correct results. The underlying principle of the 
construction of $V$ there is by the normalization $-X(\sqrt{X\dg X}\,)\1$ of 
an operator $X$, where in that case $X$ is the Wilson-Dirac operator. Another 
type of construction occurs in an example of Chiu \cite{ch01}, in which $V$
can be identified as the Cayley transform $V=-(Y-i\Id)(Y+i\Id)\1$ of a suitable 
hermitian operator $Y\,$. This, however, because bounded $Y$ do not allow to
reach the eigenvalue $-1$ of $V\,$, can provide proper topological properties 
only in the limit \cite{ke02}. Thus for the study of theoretical issues, where 
an explicit construction is desirable, up to now one has to rely on the 
normalization method. 

We note that operators $D$ satisfying the general GW relation \cite{gi82}
\be
\{\ga,D\}=2D\ga RD \mb{with} [R,D]\ne0
\label{gGW}
\ee
(where $R\dg=R$ and $[\gamma_{\mu},R]=0$) do {\it not} belong to the class. In 
fact, with $\ga$-hermiticity of $D$ and $[\ga,R]=0$ from this relation one gets
$[D,D\dg]=2D\dg[R,D]D\dg$. Thus for $[R,D]\ne0$ the Dirac operator $D$ is not 
normal, which is in contrast to what is required in \re{COM}.

In this context it is to be mentioned that for the general GW relation \re{gGW}
the analysis of the index gets rather subtle. One obtains the relation 
\be
\mbox{Tr}\Big(\ga(P_j+RQ_j)\Big)+\mbox{Tr}(\ga RD)=0 
\ee 
in which the projections $P_j$ need not to be orthogonal and where 
eigennilpotents $Q_j$ can occur if the dimensions of the respective  
algebraic and geometric eigenspaces differ. In a lenghty proof \cite{ke01a}
it has been shown that for the eigenvalue $\la_k=0$ of $D$ these dimensions 
are equal so that $Q_k=0$ and $P_k=P_k\dg$. Thus for zero eigenvalue the 
unwanted term with $Q_k$ disappears and $P_k$ gets orthogonal as needed.

\subsection{Proposal of Fujikawa}

\hspace{3mm}    
The Dirac operators proposed by Fujikawa \cite{fu00} satisfy
\be
\{\ga,\ga D\}=2a_0^{2k+1}(\ga D)^{2k+2}\;,\quad k=0,1,2,\ldots\quad.
\label{FU1}
\ee
They are normal \cite{ch00}, $D\dg D=DD\dg\,$. One may also consider them as
satisfying the general GW relation with $R=a_0^{2k+1}(DD\dg)^k\,$, where,
however, in contrast to \re{gGW} one has $[R,D]=0\,$. Using $\ga$-hermiticity 
one gets from \re{FU1}
\be
D+D\dg=2a_0^{2k+1}(D\dg D)^{k+1}\;.
\label{FU2}
\ee
Comparing this with \re{gg1}, $D+D\dg V=0\,$, it follows that
\be
V=1-2a_0^{2k+1}D(D\dg D)^k=1-2a_0^{2k+1}D(\ga D)^{2k}\;,
\label{FU3}
\ee
which obviously is $\ga$-hermitian and using \re{FU2} can be checked to be 
indeed unitary. With this it follows that \re{gg} is fulfilled so that the 
operators proposed in Ref.~\cite{fu00} are a special case of our general class.

For $D$ one obtains from \re{FU3}
\be
D=a_0\1\Bigg(\h\Bigg)^{1/(2k+1)}\ga\,\Big(\ga(1-V)\Big)^{1/(2k+1)}\;,
\label{FU4}
\ee
which is the form of Ref.~\cite{fu00}. The operator $V$ here is given 
by the normalization 
\be
V=-D\W^{(2k+1)}\Big(\sqrt{D\W^{(2k+1)\dag}D\W^{(2k+1)}}\,\Big)\1
\label{VFU}
\ee
of the generalized Wilson-Dirac operator 
\be
D\W^{(2k+1)}=\Big(\h\sum_{\mu}\gamma_{\mu}(\na{\mu}-\na{\mu}\dg)\Big)^{2k+1} 
+\Big(\frac{r}{2}\sum_{\mu}\na{\mu}\dg\na{\mu}\Big)^{2k+1} + m^{2k+1} \,,
\label{WDI}
\ee
where $(\na{\mu})_{n'n} = \delta_{n'n}^4- U_{\mu n}\delta_{n',n+\hat{\mu}}^4$
and $-2r<m\le 0\,$. For $k=0$ obviously \re{FU4} becomes the overlap Dirac 
operator of Neuberger \cite{ne98}. 

The form \re{FU4}, involving in addition operators $\ga$, is not yet the one
which according to our considerations should occur. To get the latter 
we note that because of $D\dg=-DV\dg$ we can also write \re{FU3} as
\be
V=1-2a_0^{2k+1}(-V)^{-k}D^{2k+1}\;.
\label{FU5}
\ee
With this we obtain 
\be
D=a_0\1\,\Bigg(\h(1-V)(-V)^k\Bigg)^{1/(2k+1)}\;,
\label{FU6}
\ee
where of the $(2k+1)$-th roots the one which satisfies \re{cond*o} is to be 
choosen. Thus with \re{FU6} we have indeed the form $D=F(V)$ required by the 
general concept, which appears more convenient in practice, too.

The functions $f$ of the spectral representation \re{specd} for $D\,$, in the
present case are obtained from \re{FU6} as
\be
f(\e^{i\vp})=a_0\1\,\e^{i(\vp-\pi)/2}\Big(\sin\frac{\vp}{2}\Big)^{1/(2k+1)}
\label{rs}
\ee
where of the $(2k+1)$-th roots the real one is to be taken. They give the
curves describing the location of the eigenvalues of $D\,$, which for $k>0$ 
arise as deformations of the circle for $k=0\,$. All of them meet the real 
axis at zero and at the value $f(-1)=a_0\1\,$.

\subsection{Extension of Fujikawa type}

\hspace{3mm}    
Recently an extension of the proposal of Fujikawa has been reported 
\cite{fu02}, the Dirac operators of which satisfy
\be
\{\ga,\ga D\}=2a_1D\dg D\;\Phi(a_1^2D\dg D)\;,\quad
\label{FU11}
\ee
where the operator function $\Phi$ is subject to $\Phi(X)\dg=\Phi(X)$ for 
$X\dg=X\,$. From the identity $[\ga,D\dg D]=[\{\ga,\ga D\},\ga D]$ with 
\re{FU11} since $D\dg D=(\ga D)^2$ one gets 
\be
[\ga,D\dg D]=0\;.
\label{FU15}
\ee 
With this one has $D\dg D=\ga D\dg D\ga$ and finds 
\be
[D\dg,D]=0\;.
\label{FU14}
\ee

Using $\ga$-hermiticity of $D$ to write \re{FU11} in the form 
$D+D\dg=2a_1D\dg D\;\Phi(a_1^2D\dg D)$ and comparing with the basic relation
\re{gg1}, $D+D\dg V=0\,$, it follows that
\be
V=1-2a_1D\,\Phi(a_1^2D\dg D)\;.
\label{FU12}
\ee
Because of \re{FU15} and $\Phi(a_1^2D\dg D)\dg=\Phi(a_1^2D\dg D)$ this is seen 
to be $\ga$-hermitian and using \re{FU14} and \re{FU11} it can be checked to 
be unitary. It thus turns out that the operators satisfying \re{FU11} are also
a special case of our general class.

With \re{FU12} and \re{gg1} one gets the equation 
\be
2a_1D\,\Phi(-a_1^2V\1D^2)+V=1
\label{FU13}
\ee
for $D$ and $V$, the solution of which gives $D=F(V)\,$.

\section{Realizations of subclasses}\se

\subsection{Expansion in powers of $V$}

\hspace{3mm}    
We first realize functions $D=F(V)$ of the class in terms of expansions 
$D=\sum_{\nu} c_{-\nu}V^{\nu}\,$. According to \re{cond*o} the 
coefficients $c_{\nu}$ must be real and \re{gg1} requires them to 
satisfy $c_{\nu-1}=-c_{-\nu}\,$. We thus obtain
\be
D=\sum_{\nu\ge0}c_{\nu}(V^{-\nu}-V^{\nu+1})\;,
\label{DVt}
\ee
for which $f(1)=0$ is seen to hold. Condition \re{conda} now gets 
the form
\be
f(-1)=2\sum_{\nu\ge0}(-1)^{\nu}c_{\nu}\ne0\,.
\label{condc}
\ee
Obviously \re{DN} is a special case of \re{DVt} and arises by putting 
$c_{\nu}=\rho\,\delta_{\nu0}\,$. Other choices of the coefficients are seen
not to allow GW-type relations. 

For the functions $f$ in the spectral representation of the Dirac operators 
\re{DVt} we obtain the form
\be
f(\e^{i\vp})=2\e^{i(\vp-\pi)/2}\sum_{\nu\ge0}c_{\nu}\sin(2\nu+1)\frac{\vp}{2}
\;.
\label{Dla}
\ee
The functions $\e^{i(\vp-\pi)/2} \sin(2\nu+1)\vp/2$ in \re{Dla} describes 
rosettes in the complex plane. The eigenvalues of $D$ thus reside on a closed 
curve which is given by a linear combination of rosette functions and which 
meets the real axis at zero and at the value \re{condc}. 

In case of an infinite number of terms of the expansion \re{DVt}, convergence 
properties can be studied considering the Fourier series
$\sum_{\nu\ge0}c_{\nu}\sin(2\nu+1)\frac{\vp}{2}$ in \re{Dla}.
Uniform convergence then is guaranteed by the condition
\be
\sum_{\nu\ge0}|c_{\nu}|<\infty \;.
\label{GLM}
\ee

Having $F(V)$ it remains to specify $V$. Below we show that the overlap $V\,$ 
is appropriate here, for which the explicit form $V=-D\W^{(1)}\Big(\sqrt{D
\W^{(1)\dag} D\W^{(1)}}\,\Big)\1$ with $D\W^{(1)}$ given by \re{WDI} is 
available. In principle this could also be done for other $V$ suitable in the 
GW case \re{DN}, since with respect to the tests considered they should be 
equivalent to the overlap $V\,$. 

The continuum limit of the propagator gives us further conditions on the 
coefficients $c_{\nu}\,$. To study the propagator we note that in the free 
case and infinite volume with the Fourier representation $V_{\ka'\ka}=
\tilde{V}(\ka)\delta_{\ka'\ka}^4$ on gets for the powers of $V$ with 
$\ka_{\mu}=ap_{\mu}$
\be 
\tilde{V}^{-\nu}(\ka)-\tilde{V}^{\nu+1}(\ka)\ra a\,\frac{2\nu+1}{|m|}\,i
\sum_{\mu}\gamma_{\mu}p_{\mu}\mb{for}a\ra0\;, 
\label{TV}
\ee
while at the corners of the Brillouin zone one has
\be 
\tilde{V}^{-\nu}- \tilde{V}^{\nu+1}= 2(-1)^{\nu}\;.
\label{TV0}
\ee
According to \re{DVt} and \re{TV} for the usual normalization of the continuum 
propagator
\be
\sum_{\nu\ge0}(2\nu+1)c_{\nu}\ra |m|a\1 \mb{for}a\ra0
\label{condb}
\ee
is needed, while with \re{DVt} and \re{TV0} suppression of doublers requires
\be
2\sum_{\nu\ge0}(-1)^{\nu}c_{\nu} \ra \infty \mb{for}a\ra0\;.
\label{X2}
\ee
It is seen that \re{condc} is a prerequisite for \re{X2}. In the special case 
$c_{\nu}=\rho\,\delta_{\nu0}$ \re{condb} is satisfied putting $\rho=|m|a\1\,$,
by which \re{X2} gets $2|m|a\1$ and is fulfilled, too. In the general case a 
simple way to satisfy \re{condb} and \re{X2} as well as \re{condc} and \re{GLM}
is to put
\be
c_{\nu}=|m|a\1\,\hat{c}_{\nu}
\label{X3}
\ee
and to require
\be
\sum_{\nu\ge0}(2\nu+1)\hat{c}_{\nu}=1\;,\quad 
\sum_{\nu\ge0}(-1)^{\nu}\hat{c}_{\nu}\ne0\;,\quad 
\sum_{\nu\ge0}|\hat{c}_{\nu}|<\infty \;.
\label{conddx}
\ee
Obviously also weaker conditions than this can be formulated.

With respect to topological charge and chiral anomaly nothing remains to be
shown. The reason for this is that only $V$ enters the relation for the index
\re{INN} and that the overlap $V$ is safely known to give the correct limit 
of the topological charge (see Ref.~\cite{ke01} for a proof and a discussion 
of literature). Thus the correct results are also guaranteed here.

\subsection{Construction with spectral functions}

\hspace{3mm} 
We next give a more general construction of Dirac operators of the class
using the functions $f$ of the spectral representation \re{specd} as a tool. 
We start noting that condition \re{cond+}, $f(\e^{i\vp})+f(\e^{i\vp})^*
\e^{i\vp}=0\,$, can be written as
\be
\Big(i\e^{-i\vp/2}f(\e^{i\vp})\Big)^*=i\e^{-i\vp/2}f(\e^{i\vp})\;.
\label{SY1}
\ee
This shows that $f$ is of form
\be
f(\e^{i\vp})=\e^{i(\vp-\pi)/2}\,g(\vp)\;,\quad g(\vp)\;\mbox{real}\;.
\label{SY2}
\ee
Then from condition \re{cond*}, $f(\e^{i\vp})^*=f(\e^{-i\vp})$, the 
requirement
\be 
g(\vp)=-g(-\vp)
\label{SY3}
\ee
follows. Further, with $(2\pi)$-periodicity in $\vp$ of $f(\e^{i\vp})$, the 
function $g(\vp)$ has to satisfy
\be
\quad g(\vp+2\pi)=-g(\vp)\;.
\label{SY4}
\ee
We now see that forms of $D=F(V)$ can be obtained by determining functions 
$g(\vp)$ which are real, odd, and satisfy \re{SY4}.

With the indicated requirements the basic building blocks for the construction 
of $g$ are the functions $\sin\!\frac{\nu\vp}{2}$ and $\cos\!\frac{\mu\vp}{2}$ 
with integer $\nu$ and $\mu\,$. The conditions on $g$ are fulfilled by 
$\sin\!\frac{\nu\vp}{2}$ with odd $\nu$ and $\sin\!\frac{\nu\vp}{2}\cos\!
\frac{\mu\vp}{2}$ with even $\nu$ and odd $\mu$, where the latter form, 
however, can be expressed by terms of former type. Further, after 
multiplication with a function of the $\cos\frac{\mu\vp}{2}$ with even integers
$\mu\,$ one has still the required properties. Thus, with the notation
\be
s_{\nu}=\sin(2\nu+1)\frac{\vp}{2}\;,\quad t_{\mu}=\cos\mu\vp\;,\qquad\nu,\mu
\;{\rm integer}\;,
\label{GE1}
\ee
we arrive at the form
\be
g=\sum_{\nu}s_{\nu}w_{\nu}(t_1,t_2,\ldots) 
\label{GE1a}
\ee 
where the $w_{\nu}$ are real functions. Because of the identity
\be
s_{\nu}=s_0 \Big(1+2\sum_{\mu=1}^{\nu}t_{\mu}\Big)
\label{GE2}
\ee
\re{GE1a} can be simplified to 
\be
g=s_0\,w(t_1,t_2,\ldots)
\label{GE2a}
\ee 
with a real function $w\,$. Further, since the $t_{\nu}$ are given
by polynomials of $t_1\,$,
\be
t_{2\mu}=d_{2\mu}t_1^{2\mu}+d_{2\mu-2}t_1^{2\mu-2}+\ldots+d_0\;,\quad
t_{2\mu+1}=d_{2\mu+1}t_1^{2\mu+1}+d_{2\mu-1}t_1^{2\mu-1}+\ldots+d_1t_1\;,
\label{GE3}
\ee
\re{GE1a} can be cast into the still simpler form 
\be
g=s_0\,w(t_1)\;.
\label{GE4}
\ee
Condition \re{conda}, $f(-1)=g(\pi)\ne0\,$, in case of \re{GE4} is seen to 
require 
\be
w(-1)\ne0\;.
\label{GE4a}
\ee
Further, $w(t_1)$ should, of course, be finite in the region of interest.

The form \re{GE4} essentially extends the one found in Subsection 4.1 to the 
case where no series expansion in $V$ is available. To see this one has to 
note that there according to \re{Dla} we have
\be
g=2\sum_{\nu\ge0}c_{\nu}s_{\nu}=s_0\;\Big(b_0/2+\sum_{\mu\ge1}b_{\nu}
t_{\nu}\Big)=s_0\;w(t_1)\;,
\label{GE13}
\ee
where the second form follows inserting \re{GE2} into the first one and 
introducing coefficients $b_{\mu}=2\sum_{\nu\ge\mu}c_{\nu}$. Then using 
\re{GE3} from the second one the last one is obtained, which is \re{GE4} 
specialized to the case where $w(t_1)$ is given by a power series.

The operator $D$ for the construction \re{SY2} with \re{GE4},  
\be
f(\e^{i\vp})=\e^{i(\vp-\pi)/2}s_0\,w(t_1)=\textstyle{\h}(1-\e^{i\vp})
w(\cos\vp)\;,
\label{GAi}
\ee
inserting \re{GAi} into \re{specd} becomes
\be
D=\textstyle{\h}(\Id-V)\;W\Big(\textstyle{\h}(V+V\dg)\Big)\;,
\label{GAO1}
\ee
where the operator function $W$ has to satisfy
\be
W(X)\dg=W(X\dg)\;,\quad W(-P)\ne0\;\;{\rm for}\;\;P=P^2=P\dg>0\;.
\label{GA1i}
\ee

To check the continuum limit it remains to consider the operator 
$W\Big(\h(V+V\dg)\Big)$ which as compared to the GW case \re{DN} replaces the 
constant $2\rho\,$. This can again be done with the overlap $V\,$. For 
$\h(V+V\dg)$ in the free case and infinite volume we get with 
$\ka_{\mu}=ap_{\mu}$
\be 
\textstyle{\h}\Big(\tilde{V}(\ka)+\tilde{V}\dg(\ka)\Big)\ra1\mb{for}a\ra0\;, 
\label{TV1}
\ee
while at the corners of the Brillouin zone we have
\be
\textstyle{\h}\Big(\tilde{V}+\tilde{V}\dg\Big)=-1\;.
\label{TV2}
\ee
Thus the requirements \re{condb} and \re{X2} generalize to
\be
\tilde{W}(1)\ra 2|m|a\1\;,\quad \tilde{W}(-1)\ra\infty \mb{for}a\ra0\;.
\label{TV3}
\ee
These conditions can be satisfied by putting
\be
\tilde{W}=2|m|a\1\hat{W}\;,\quad \hat{W}(1)=1\;,\quad \hat{W}(-1)\ne0\;,
\label{TV4}
\ee
though again also weaker conditions can be formulated.

With respect to the meaning of these conditions we note that the eigenvalues of
$V$ in the free case are those of $\tilde{V}\,$,
\be
\e^{i\vp}=-(\tau\pm i\sqrt{s^2})/\sqrt{\tau^2+s^2}\;,
\label{TU1}
\ee
where
\be
s^2=\sum_{\mu}\sin\ka_{\mu}^2\;,\quad \tau=m+r\sum_{\mu}(1-\cos\ka_{\mu})\;,
\quad -2r<m<0\;.
\ee
The real eigenvalues in \re{TU1} obviously occur for $\ka_{\mu}=0,\pi\,$ and 
one gets $+1$ if all $\ka_{\mu}=0$ and $-1$ at each corner of the Brillouin 
zone. Noting that
\be
\cos\vp=\textstyle{\h}\Big(\tilde{V}(\ka)+\tilde{V}\dg(\ka)\Big)=
-\tau/\sqrt{\tau^2+s^2}\;, 
\ee
the conditions on $\tilde{W}(1)$ and $\tilde{W}(-1)$ are seen to be related to 
the behavior of $V$ at its eigenvalues $+1$ and $-1\,$, repectively.

\subsection{Extension of spectral construction}

\hspace{3mm} 
To extend the construction of the preceeding Section we note that given a 
function $g$ with the required properties, then $h(g)$ is again a function 
with such properties provided that $h$ is odd and real, 
\be
h(-x)=-h(x)\;,\quad h(x)^*=h(x)\;\;\mbox{for real}\;\;x\;.
\label{GE5}
\ee
With this the form \re{GE4} generalizes to
\be
g=h\Big(s_0\,w(t_1)\Big)\;.
\label{GE6}
\ee
Now condition \re{GE4a} extends to  
\be
h\Big(w(-1)\Big)\ne0\;,
\label{GE6a}
\ee
which is satisfied anticipating the requirement of strict monotony of $h$ 
implied by \re{MON}. 

Looking for functions $h$ an important observation is that because of the 
identity
\be
s_0^{2k+1}=s_0\,\sum_{\nu=0}^k\Big(\begin{array}{c}\scriptstyle 2k+1\\
\scriptstyle\nu\\\end{array}\Big) (-1)^{\nu}\Big(1+2\sum_{\mu=1}^{k-\nu}
t_{\nu}\Big)\;,\quad k=0,1,2,\ldots\;,
\label{GE7}
\ee 
\re{GE6} reduces to the form \re{GE4} if $h(x)$ is an integer power, is a 
polynomial, or allows an expansion in powers of $x$. This greatly restricts 
nontrivial choices of $h(x)$, i.e.~ones by which \re{GE6} gives something 
beyond \re{GE4}.

An example of nontrivial choices in this sense, which can be expressed in an 
elementary way, is
\be 
h(x)=x^{1/(2k+1)}\;,\;\; k=1,2,\ldots\;
\label{GE9}
\ee
where the real one of the $(2k+1)$-th roots is to be taken. One should note
that with this one has a different nontrivial choice for each $k\,$.

The nontrivial choices of $h$ actually are equivalence classes, i.e.~equivalent
ones are not to be counted as different. For example, to \re{GE9} the form
$h(x)=x^{(2k+1)}r(x)$ with $r(-x)=r(x)$ is equivalent, because with $r(x)=
r(|x|)$ and $|s_0|=\sqrt{\h(1-t_1)}$ it gives the same form in \re{GE6} as 
\re{GE9}. Also forming in addition odd powers because of \re{GE7} gives 
nothing new.

The number of nontrivial choices of $h(x)$ which can be explicitly expressed 
in terms of elementary functions appears to be limited. We give a further
explicit example in Subsection 4.7 and also point out strategies for more 
general constructions there.

The operator $D$ for the construction \re{SY2} with \re{GE6},  
\be
f(\e^{i\vp})=\e^{i(\vp-\pi)/2}\,h\Big(s_0\,w(t_0)\Big)\;,
\label{GB}
\ee
inserting \re{GB} into \re{specd} becomes
\be
D=-iV^{\h}\,H\Bigg(\frac{1}{2i}(V^{\h}-V^{-\h})\;W\Big(\textstyle{\h}(V+V\dg)
\Big)\Bigg)\;,
\label{GAO}
\ee
where $V^{\h}$ is defined with $+\e^{i\vp_k/2}$ in its spectral representation,
$W$ has to satisfy \re{GA1i} and for the function $H$ one needs
\be
H(-X)=-H(X)\;,\quad H(X)\dg=H(X)\;\;\mbox{for}\;\;X\dg=X\;.
\label{GB1}
\ee
Eq.~\re{GAO} demonstrates the drastic increase of possible forms of Dirac
operators which occurs as compared to the GW case.

\subsection{Example $h(x)=x^{1/(2k+1)}$}

\hspace{3mm} 
To check the continuum limit for a particular choice of a nontrivial $h$, 
we use \re{GE9} as an example, where $h$ is conveniently
represented in terms elementary functions. With it \re{SY2} becomes
\be
f(\e^{i\vp})=\e^{i(\vp-\pi)/2}\Big(s_0\,w(t_1)\Big)^{1/(2k+1)}=
\textstyle{\h}(1-\e^{i\vp})^{1/(2k+1)}w(\cos\vp)^{1/(2k+1)}\;.
\label{GAj}
\ee
The operator $D$ inserting 
\re{GAj} into \re{specd} and putting $W^{1/(2k+1)}=W_k$ becomes
\be
D=\Bigg(\h(1-V)(-V)^k\Bigg)^{1/(2k+1)}W_k\Big(\textstyle{\h}
(V+V\dg)\Big)\;,
\label{GAO2}
\ee
where of the $(2k+1)$-th roots the one which satisfies \re{cond*o} is to be 
choosen and where $W$ is subject to \re{GA1i}. 

Obviously \re{GAO2} generalizes our form \re{FU6} of Fujikawa's proposal 
replacing the constant $a_0\1$ there by a function $W_k\Big(\h(V+V\dg)\Big)\,$.
Using $V$ given by \re{VFU} with \re{WDI}, in the free case and 
infinite volume one gets with $\ka_{\mu}=ap_{\mu}$ 
\be 
\tilde{V}(\ka)\ra 1-\Bigg(\frac{a}{|m|}\Bigg)^{2k+1}\,\Big(i\sum_{\mu}
\gamma_{\mu} p_{\mu}\Big)^{2k+1}\mb{for}a\ra0\;, 
\label{TVF}
\ee
while at the corners of the Brillouin zone one has $\tilde{V}=-1\,$.
The requirements \re{TV3} thus generalize to
\be
\tilde{W}_k(1)\ra 2^{1/(2k+1)}(-1)^k|m|a\1\;,\quad\tilde{W_k}(-1)\ra\infty\;
\mb{for}a\ra0\;.
\label{TVFa}
\ee
This can be satisfied by putting
\be
\tilde{W_k}=2^{1/(2k+1)}(-1)^k|m|a\1\;\hat{W_k}\;\;,\quad \hat{W_k}(1)=1\;,
\quad \hat{W_k}(-1)\ne0\;,
\label{TVFd}
\ee
or by some weaker condition. In the special case where $W_k$ is replaced by 
the constant $a_0\1$ according to \re{TVFd} one gets$\,$\footnote{
Starting from \re{FU4} as in literature the factor $(-1)^k\,$ also arises by 
properly calculating $\ga(\sum_{\mu}\gamma_{\mu} s_{\mu})^{-(2k+1)}=(-1)^k
(\ga\sum_{\mu} \gamma_{\mu}s_{\mu})^{-(2k+1)}$ before taking the $(2k+1)$-th 
root.}
\be
a_0\1=2^{1/(2k+1)}(-1)^k|m|\;a\1\;.
\label{TVFc}
\ee

The eigenvalues of $V$ in the free case here are 
\be
\e^{i\vp}=-\Big(\tau_k\pm i\sqrt{(s^2)^{2k+1}}\Big)/\sqrt{\tau_k^2+(s^2)
^{2k+1}}\;,
\label{TU1j}
\ee
with
\be
s^2=\sum_{\mu}\sin\ka_{\mu}^2\;,\quad \tau_k=m^{2k+1}+\Big(r\sum_{\mu}(1-
\cos\ka_{\mu})\Big)^{2k+1}\;, \quad -2r<m<0\;.
\ee
In \re{TU1j} the eigenvalues also get $+1$ if all $\ka_{\mu}=0$ and $-1$ at 
each corner of the Brillouin zone. Noting that 
\be
\cos\vp=\textstyle{\h}\Big(\tilde{V}(\ka)+\tilde{V}\dg(\ka)\Big)=
-\tau_k/\sqrt{\tau_k^2+(s^2)^{2k+1}}\;, 
\ee
the conditions on $\tilde{W}(1)$ and $\tilde{W}(-1)$ are again seen to be
related to the behavior of $V$ at its eigenvalues $+1$ and $-1\,$, repectively.

\subsection{Comparison of conditions on $H$ and $W$}

\hspace{3mm} 
The example of the Dirac operator \re{GAO2} in Subsection 4.4 
contains the operator related to the GW relation \re{GW}, Fujikawa's operator
\re{FU4}, and our constructions \re{DVt} and \re{GAO1} (where \re{DVt} has 
been seen to be  the expansion version of \re{GAO1}$\,$) as special cases. Thus
it allows us to compare the features of the continuum limit for $V$ from 
\re{VFU} with \re{WDI} (the case with more general $V$ being considered in 
Subsection 4.6$\,$).

The behavior of $V$ is essentially determined by the choice of $h(x)$, which
for $h(x)=x^{1/(2k+1)}$ is given by \re{TVF},
\be 
\tilde{V}(\ka)\ra 1-\Bigg(\frac{a}{|m|}\Bigg)^{2k+1}\,\Big(i\sum_{\mu}
\gamma_{\mu} p_{\mu}\Big)^{2k+1}\mb{for}a\ra0\;. 
\label{TVF1}
\ee
This makes the form for $k>0$ as well as for $k=0$ explicit and 
illustrates the impact of the function $H$ (which is more generally seen in
\re{TVF7}$\,$).

With respect to the condition on $W$, which leads to the usual normalization
of the continuum propagator, one has from \re{TVFa}, 
\be
\tilde{W_k}(1)\ra 2^{1/(2k+1)}(-1)^k|m|a\1\; \mb{for}a\ra0\;,
\label{TVFb}
\ee
which as well as the suppression of doublers is guaranteed by \re{TVFd}.
In Fujikawa's case $W_k=a_0\1$ is a constant, \re{TVFb} becomes
$a_0\1=2^{1/(2k+1)}(-1)^k|m|a\1$ and \re{TVFd} is trivially satisfied. In 
case of the GW relation \re{GW}, where $W=a_0\1=2\rho$, this reduces to 
$\rho=|m|a\1$. Thus the straightforward connection to the special cases
becomes obvious and it is seen that the essential requirements are the
same ones in all cases.

\subsection{Related sets of $V$}

\hspace{3mm} 
Clearly the determination of the functions $F(V)$ and the selection of the
operators $V$ are distinct issues. Nevertheless they are not completely 
independent. For the choice $h(x)=x$ the set of $V$ is appropriate which is
known to be suitable already in the GW case. However, for $h(x)=x^{1/(2k+1)}$ 
for different $k$ different $V$ are needed. This suggests that 
the {\it nontrivial} choices of $h$ relate to distinct sets of $V$.

Indeed, what occurs in the above examples extends to other functions $h$
for which a strictly monotonous inverse function $\eta$ exists, 
\be
h\Big(\eta(x)\Big)=x\;,\qquad \eta(y)>\eta(x)\;\;{\rm for}\;\;y>x\;.
\label{MON}
\ee
In terms of operators one then has a corresponding hermitian function $E$ with 
$H(E(X))=X\,$. This allows us to introduce the general normalization-type 
definition of $V\,$, 
\be
V=-D\W^{(\eta)}\Big(\sqrt{D\W^{(\eta)\dag}D\W^{(\eta)}}\,\Big)\1
\label{Vg}
\ee
with
\be
D\W^{(\eta)}=iE\Big(\frac{1}{2i}\sum_{\mu}\gamma_{\mu}(\na{\mu}-\na{\mu}\dg)
\Big) +E\Big(\frac{r}{2}\sum_{\mu}\na{\mu}\dg\na{\mu}\Big) + E(m\Id) \;,
\label{VgD}
\ee
where the function $E$ of the operators (which are hermitian) is well-defined. 
This definition of $V$ is to be inserted into the general expression for $D\,$, 
\be
D=-iV^{\h}a\1H\Bigg(\frac{1}{2i}(V^{\h}-V^{-\h})\;W\Big(\textstyle{\h}(V+V\dg)
\Big)\Bigg) \;,
\label{GAOa}
\ee
which is \re{GAO} up to a factor $a\1$. In the free case for this $V$ at 
the corners of the Brillouin zone one gets $\tilde{V}=-1$ and at zero
\be 
\tilde{V}\ra 1-\frac{i}{|\eta(m)|}\,\tilde{E}\Big(a
\sum_{\mu}\gamma_{\mu}p_{\mu}\Big)\mb{for}a\ra0\;.
\label{TVF7}
\ee
Requiring $\tilde{W}(-1)\ne0$, because of the monotony of $E(X)$ doublers are 
suppressed for $-2r<m<0$ as usual. Since $H(E(X))=X$, putting 
$\tilde{W}(1)=2|\eta(m)|$ the correct limit of the propagator is obtained in 
this quite general way. 

We note that the strict use of hermitian functions of hermitian operators in 
\re{Vg} with \re{VgD} is what makes it applicable in a general way. In contrast
to this the formulation of \re{VFU} with \re{WDI} appears restricted to the 
choice $h(x)=x^{1/(2k+1)}$. Treating the latter choice with the more 
satisfactory form \re{Vg} with \re{VgD}, instead of \re{TVF} one gets 
\be 
\tilde{V}(\ka)\ra 1-i\Bigg(\frac{a}{|m|}\Bigg)^{2k+1}\,\Big(\sum_{\mu}
\gamma_{\mu} p_{\mu}\Big)^{2k+1}\mb{for}a\ra0\;, 
\label{TVF2}
\ee
with different position of the factor $i\,$. On the other hand, instead of 
the first relation in \re{TVFa} one then has
\be
\tilde{W}(1)= 2|m|^{2k+1}\;, \quad \tilde{W}_k(1)=2^{1/(2k+1)}|m|\;,
\label{TVFe}
\ee 
in which the factors $(-1)^k$ and $a\1$ of \re{TVFa} do not occur. The latter
factor is now incorporated in the definition \re{GAOa} of $D$ while the former 
one is accounted for by the different position of $i$ in \re{TVF2}. Thus for 
the choice $h(x)=x^{1/(2k+1)}$, for which both formulations of $V$ are 
applicable, one gets only minor differences in the representations.

\subsection{Chirality of $D\1$}

\hspace{3mm} 
If the inverse $D\1$ exists one may multiply \re{GW} from both sides with
it, which gives
\be
\{D\1,\ga\}=\rho\1\;.
\ee
This relation indicates that the propagator is chiral up to a local contact 
term, which is desirable to get the Ward identities of chiral symmetry in the
continuum limit. In case of Fujikawa's operator this relation becomes
\be
\{D\1,\ga\}=2a_0^{2k+1}(D\ga)^{2k}
           =a_0\Big(2(1-V)^{2k}(-V)^{-k}\Big)^{1/(2k+1)}\;,
\label{LO0}
\ee
so that the locality of interest follows from that of $D$. 
For the example of the Dirac operator \re{GAO2}, which contains these 
operators as well as  our constructions \re{DVt} and \re{GAO1} as special 
cases, we obtain
\be
\{D\1,\ga\}=\Big(2(1-V)^{2k}(-V)^{-k}\Big)^{1/(2k+1)}\;W_k\Big(\textstyle{\h}
(V+V\dg)\Big)\1 \;.
\label{LO1}
\ee
Since this differs from \re{LO0} only in that $a_0$ is replaced by 
the function $W_k\Big(\textstyle{\h}(V+V\dg)\Big)\1$, the additional
requirement is that the properties of $W$ should be such that the product 
on the r.h.s.~of \re{LO1} remains local.

The above strategy to relate the locality of $\{D\1,\ga\}$ to that of $D$ 
relies on the introduction of an inverse function to $D=F(V)$,
\be
V=\Id-I(D) \;,
\ee
with which from the basic relation \re{gg} one generally obtains
\be
\{D\1,\ga\}=\ga I(D)D\1\;.
\ee
In the special case of \re{LO0} according to \re{FU3} one has $I(D)=
2a_0^{2k+1}(DD\dg)^kD$ so that $I(D)D\1=2a_0^{2k+1}(DD\dg)^k$. 
Analogously in the general case it is seen that $I(D)$ must be such that 
locality of $D$ implies that of $I(D)D\1$. 

To study this in more detail we consider the subclass of Dirac operators for
which $I(D)D\1=(\Id-V)D\1$ is a polynomial of $DD\dg$,
\be
I(D)D\1=\sum_{\nu=0}^N\C_{\nu}(DD\dg)^{\nu}\;,
\label{ID1}
\ee
with real coefficients $\C_{\nu}\,$. In terms of the spectral functions
this means that
\be
(1-\e^{i\vp})f(\e^{i\vp})\1=\sum_{\nu=0}^N\C_{\nu}\Big(f(\e^{i\vp})
f(\e^{i\vp})^*\Big)^{\nu}\;,
\ee
which inserting \re{SY2}, 
$f(\e^{i\vp})=\e^{i(\vp-\pi)/2}\,g(\vp)$, becomes
\be
2s_0(\vp)=\sum_{\nu=0}^N\C_{\nu}\,g(\vp)^{2\nu+1}\;,\qquad s_0=\sin\frac{\vp}
{2}\;.
\label{ID2}
\ee
To obtain $D$ by inserting $f$ into \re{specd}, we have to determine $g$,
which according to \re{ID2} requires to solve the algebraic equation 
\be
\sum_{\nu=0}^N\C_{\nu}\,g^{2\nu+1}-2s_0=0\;.
\label{ALG}
\ee  
The solution of \re{ALG} is trivial if only one of the coefficients $\C_{\nu}$ 
is different from zero, which gives Fujikawa's proposal. 

In order to derive a more general explicit solution we consider the case where 
the two coefficients $\C_0$ and $\C_1$ are nonzero. We then have the cubic 
equation
\be  
g^3+3pg+2q=0\;,\qquad p=\frac{\C_0}{3\C_1}\;,\quad q=-\frac{s_0}{\C_1}\;.
\label{ALG3}
\ee
Requiring $\C_0>0$ and $\C_1>0$ it follows that $p^3+q^2>0$, which implies 
that one gets the real solution
\be
g=\sqrt[3]{-q+\sqrt{q^2+p^3}}+\sqrt[3]{-q-\sqrt{q^2+p^3}}\;.
\ee
Thus in more detail we arrive at
\be
g=h\Big(\frac{s_0}{\C_1}\Big)=\sqrt[3]{\frac{s_0}{\C_1}+\sqrt{\Big(\frac{s_0}
{\C_1}\Big)^2+\Big(\frac{\C_0}{3\C_1}\Big)^3}}+\sqrt[3]{\frac{s_0}{\C_1}-
\sqrt{\Big(\frac{s_0}{\C_1}\Big)^2+(\frac{\C_0}{3\C_1}\Big)^3}}\;,
\label{EXA}
\ee
where the new nontrivial function $h$ has the required properties of being odd 
and strictly monotonous. The associated inverse function according to \re{ALG3}
is $\eta(g)=\h(g^3+\frac{\C_0}{\C_1}g)\,$. In the related operator expression 
\re{GAOa} for $D$ now $H$ is of form \re{EXA} with $s_0$ replaced 
by
$\frac{1}{2i}(V^{\h}-V^{-\h})$ and $W$ is constant, $W=1/\C_1$. The condition
for the usual normalization of the propagator in the continuum limit thus 
becomes $1/\C_1=2|\eta(m)|$ or $\C_1|m|^3+\C_0|m|=1$.

It is now seen how one could proceed to still more general cases. Firstly,
one could solve \re{ALG} for more coefficients. Secondly, one could consider 
more complicated expressions on the r.h.s.~of \re{ID1}.

It should, furthermore, be noted that instead of the indicated strategy in 
many cases it may be more convenient to analyze directly 
\be
\{D\1,\ga\}=-ia(V^{\h}-V^{-\h})H\Bigg(\frac{1}{2i}(V^{\h}-V^{-\h})\;
W\Big(\textstyle{\h}(V+V\dg)\Big)\Bigg)\1 \;,
\label{LOa}
\ee
which follows from the general form \re{GAOa} of $D$. Generally the desired
locality obviously puts some additional conditions on the choices of the 
functions $H$ and $W$, however, leaving still much freedom.

\section{Weyl operators and chiral gauge theories}\se

\hspace{3mm}    
The chiral projection operators implicit in the overlap formalism of Narayanan 
and Neuberger \cite{na93} and used in the formulation of L\"uscher \cite{lu98}
in our notation read 
\be
P_{\pm}=P_{\pm}\dg=\h(1\pm\ga)\Id\;,\quad \tilde{P}_{\pm}=\tilde{P}_{\pm}\dg=
\h(1\pm \ga V)\Id\;.
\label{PR} 
\ee
Obviously only $\ga$ and $V$ are involved in them and we can start with them 
only requiring generalized chiral symmetry. From condition \re{gg} we get the 
identity $D=\h(D-\ga D\ga V)$ and inserting 
$\ga=P_+-P_-$ and $\ga V=\tilde{P}_+ -\tilde{P}_-$ into it we obtain
\be
D=P_+D\tilde{P}_-+P_-D\tilde{P}_+\;.
\label{DP}
\ee
With this we have the relations 
\be
P_{\pm}D\tilde{P}_{\mp}= D\tilde{P}_{\mp}= P_{\pm}D \;,
\label{PD}
\ee 
which generalize the expressions for the Weyl operators in terms of the Dirac 
operator familiar in continuum theory. 

With respect to possible forms of \re{PD} one should be aware of the fact that
the relations 
\be
P_{\pm}\ga=\pm\ga\Id\;,\quad \ga V\tilde{P}_{\mp}=\mp\tilde{P}_{\mp}\;, 
\label{PgV}
\ee
allow to absorb parts of $D\,$. In the special case of the Dirac operator 
\re{DN}, with \re{PgV} one gets $P_+\rho(1-V)\tilde{P}_-=2\rho P_+
\tilde{P}_-\, $, which relates the different forms of the chiral determinant 
in Ref.~\cite{lu98} and in Ref.~\cite{na93}, commented on in Refs.~\cite{ne99,
go00}. Considering the general class of operators $D$ here, we have to observe
that \re{PD} is the generally valid form and that modifications by \re{PgV} 
depend on the particular choice of $D\,$. Furthermore, we also note that for 
any operator $C$ satisfying $\ga C-C\ga V=0$ one gets $P_{\pm}D\tilde{P}_{\mp}=
P_{\pm}(D+C) \tilde{P}_{\mp}\,$, which provides another possibility to modify 
\re{PD}. 

For the numbers of the degrees of freedom $\Tr\,P_+$ and $\Tr\,\tilde{P}_-$ 
of the Weyl fermions in $P_+D\tilde{P}_-$ one gets from \re{PR}
\be
\Tr\,P_+-\Tr\,\tilde{P}_-=\h\Tr(\ga V)\;,
\label{INNw}
\ee
which obviously depends only on $V$ for the whole class and agrees with the 
result \re{INN} for the index of the Dirac operator $D$. 

The degrees of freedom are exhibited in more detail representing the 
projections by
\be
P_+=\sum_ju_ju_j\dg\;,\quad\tilde{P}_-=\sum_k\tilde{u}_k\tilde{u}_k\dg\;,\quad  
u_i\dg u_j=\delta_{ij}\;,\quad\tilde{u}_k\dg\tilde{u}_l=\delta_{kl}\;,
\label{uu}
\ee 
with basis vectors relating the degrees of freedom to the representation 
in full space. Associating Grassmann variables $\bar{\chi_j}$ and $\chi_k$ 
to the degrees of freedom, the fermion field variables then get
\be
\bar{\psi}=\sum_j\bar{\chi_j}u_j\dg\;,\quad \psi=\sum_k\tilde{u}_k\chi_k\;.
\ee

With this for $\Tr\,\tilde{P}_-=\Tr\,P_+$ correlation functions are given by
\ba
\int\prod_l(\di\bar{\chi_l}\di\chi_l)\;\exp{(-\bar{\psi}D\psi)}\;
\psi_{n({\sy 1})}\bar{\psi}_{n(r_1)}\psi_{n({\sy 2})}\bar{\psi}_{n(r_2)}\ldots
\psi_{n({\sy f})}\bar{\psi}_{n(r_f)}=\qquad\qquad\\\sum_{s_1,\ldots,s_f}
\epsilon_{s_1s_2\ldots s_f}\; (\tilde{P}_-D\1 P_+)_{n({s_1})n(r_1)} 
\ldots(\tilde{P}_-D\1 P_+)_{n({s_f})n(r_f)}\;\;\det M\quad
\ea
where the matrix $M$ occurring in the chiral determinant is
\be
M_{jk}=u_j\dg D\tilde{u}_k\;.
\label{MM}
\ee 

We thus have arrived at the usual basic relations of chiral gauge theories
using only generalized chiral symmetry and {\it not} the GW relation.

\section{Conclusions}\se

\hspace{3mm}    
We have shown that generalized chiral symmetry provides the general principle
for the description of massless fermions on the lattice and for the 
formulation of chiral gauge theories there, while the GW relation and its
generalization by Fujikawa occur only as special cases in this context. It has
turned out that this principle gives a much larger class of operators than 
one has in the GW and Fujikawa cases. We have given a general construction of
these operators and an associated realization of the basic unitary operator
$V$ involved. The index of the Dirac operator has been found to depend only on
$V\,$.

\section*{Acknowledgement}

\hspace{3mm}
I wish to thank Michael M\"uller-Preussker and his group for their kind 
hospitality.


\begin{thebibliography}{99}

\bibitem{gi82}  P.H. Ginsparg and K.G. Wilson, 
              Phys. Rev. D 25 (1982) 2649. 
\bibitem{lu98i}  M. L\"uscher, 
              Phys. Lett. B 428 (1998) 342. 
\bibitem{lu98}  M. L\"uscher,
              Nucl. Phys. B 549 (1999) 295; 
              Nucl. Phys. B 568 (2000) 162. 
\bibitem{fu00}  K. Fujikawa,
              Nucl. Phys. B 589 (2000) 487. 
\bibitem{fu02}  K. Fujikawa, M. Ishibashi and H. Suzuki
              Phys.Lett. B 538 (2002) 197. 
\bibitem{ch98} T.-W. Chiu, 
              Phys. Rev. D 58 (1998) 074511. 
\bibitem{ha98} P. Hasenfratz, V. Laliena, F. Niedermayer,
              Phys. Lett. B 427 (1998) 125. 
\bibitem{na93}  R. Narayanan and H. Neuberger, 
              Phys. Rev. Lett. 71, (1993) 3251;
              Nucl. Phys. B 412 (1994) 574; 
              Nucl. Phys. B 443 (1995) 305. 
\bibitem{ke02}  W. Kerler, hep-lat/0202015, to appear in JHEP.
\bibitem{at68}M.F. Atiyah and I.M. Singer, Ann. of Math. 87 (1968) 546,
                Section 5.
\bibitem{ne98}  H. Neuberger,
              Phys. Lett. B 417 (1998) 141; 
              Phys. Lett. B 427 (1998) 353. 
\bibitem{ch01} T.-W. Chiu, 
              Phys. Lett. B 521 (2001) 429. 
\bibitem{ke01a} W. Kerler,
              Phys. Lett. B 510 (2001) 325. 
\bibitem{ch00}  T.-W. Chiu, 
              Nucl. Phys. B 588 (2000) 400. 
\bibitem{ke01}  W. Kerler,
              Int. J. Mod. Phys. A 16 (2001) 3117. 
\bibitem{ne99}  H. Neuberger,
              in \emph{Lattice Fermions and Structure of the Vacuum},
              V. Mitrjushkin and G. Schierholz (eds.), (Kluwer Academic 
              Publishers, 2000) p. 113.
\bibitem{go00}  M. Golterman, 
              Nucl. Phys. Proc. Suppl. 94 (2001) 189.
\end{thebibliography}
\end{document}